\documentclass[12pt]{elsarticle}
\makeatletter
\def\ps@pprintTitle{%
  \let\@oddhead\@empty
  \let\@evenhead\@empty
  \let\@oddfoot\@empty
  \let\@evenfoot\@empty}
\makeatother
\usepackage{amssymb}
\usepackage{multirow}
\usepackage{makecell}
\usepackage[normalem]{ulem}
\usepackage{xcolor}
\usepackage{amsmath,amsfonts}
\usepackage{algorithmic}
\usepackage{algorithm}
\usepackage{graphicx}
\usepackage{url}

\begin{document}

\begin{frontmatter}



\title{Multi-Resolution Model Fusion for Accelerating the Convolutional Neural Network Training}

\author[1]{Kewei Wang\corref{cor1}}
\ead{kwf5687@ece.northwestern.edu}
\author[1]{Claire Songhyun Lee}
\author[3]{Sunwoo Lee}
\author[1]{Vishu Gupta}
\author[2]{Jan Balewski}
\author[2]{Alex Sim}
\author[2]{Peter Nugent}
\author[1]{Ankit Agrawal}
\author[1]{Alok Choudhary}
\author[2]{Kesheng Wu}
\author[1]{Wei-keng Liao}

\cortext[cor1]{Corresponding author}

\affiliation[1]{organization={Northwestern University},
addressline={2145 Sheridan Rd}, 
postcode={60208}, 
city={Evanston},
state={IL},
country={USA}}
\affiliation[2]{organization={Lawrence Berkeley National Laboratory},
addressline={1 Cyclotron Rd},
postcode={94720},
city={Berkeley, CA},
country={USA}}
\affiliation[3]{organization={Inha University},
addressline={100 Inha-ro, Michuhol-gu},
postcode={22212},
city={Inha},
country={South Korea}}



\begin{abstract}
Neural networks are rapidly gaining popularity in scientific research, but training the models is often very time-consuming.
Particularly when the training data samples are large high-dimensional arrays, efficient training methodologies that can reduce the computational costs are crucial.
To reduce the training cost, we propose a Multi-Resolution Model Fusion (MRMF) method that combines models trained on reduced-resolution data and then refined with data in the original resolution.
We demonstrate that these reduced-resolution models and datasets could be generated quickly.
More importantly, the proposed approach reduces the training time by speeding up the model convergence in each fusion stage before switching to the final stage of finetuning with data in its original resolution.
This strategy ensures the final model retains high-resolution insights while benefiting from the computational efficiency of lower-resolution training.
Our experiment results demonstrate that the multi-resolution model fusion method can significantly reduce end-to-end training time while maintaining the same model accuracy.
Evaluated using two real-world scientific applications, CosmoFlow and Neuron Inverter, the proposed method improves the training time by up to 47\% and 44\%, respectively, as compared to the original resolution training, while the model accuracy is not affected.
\end{abstract}




\end{frontmatter}

\section{Introduction}
Deep neural networks (DNNs) are increasingly instrumental in diverse scientific domains, such as cosmology \cite{mathuriya2018cosmoflow}, material science \cite{agrawal2019deep, jha2019enhancing, mao2023deep}, neuroscience \cite{Ben-Shalom727974}, etc.
The large volume of data generated from scientific applications
often require the use of high-performance computing systems for effective parallel DNN training.
Given the significant amount of resources involved, optimizing each facet of this training process is crucial.
Within the paradigm of a deep learning framework, the large-scale training of DNNs involves three critical components: computation, communication among different computing elements, and I/O operations \cite{10.1145/3458817.3476181}.
Whereas a majority of work pertaining to large-scale DNN training focuses on optimizing the cost of I/O and communication \cite{10.1145/3458817.3476181, zhang2020efficient, 9680272, yang2019accelerating, 9555961, renggli2019sparcml}, the evolving complexity and scale of DNN computations consumes a significant amount of time and thus demands attention. 
For instance, data movement and communication latency can be hidden by compute-intensive GPU workloads \cite{ibrahim2021architectural} when training complex models like DeepCAM \cite{kurth2018exascale} from the MLPerf HPC benchmark \cite{farrell2021mlperf, mattson2020mlperf}.
Moreover, as computational capabilities expand and dataset sizes exponentially increase up to billions of samples, the challenge of managing high computational costs becomes more critical.

Reducing computational costs can present an acute challenge for scientific applications due to its usage of high-precision calculations and large-scale simulation data.
To mitigate this computational burden, many researchers have explored methods such as mixed precision training \cite{micikevicius2017mixed} and sparse training \cite{frankle2018lottery}.
However, while effective in reducing computational costs, these methods potentially affect the final model accuracy due to reduced precision or iterative pruning \cite{micikevicius2017mixed}.
Beyond this, algorithmic optimization methods such as Adam \cite{kingma2014adam} and AdaGrad \cite{duchi2011adaptive} have been proposed to reduce the number of iterations required to converge.
Building upon these developments, our work leverages the inherent continuity of scientific data to approach this issue in an orthogonal way.

Our previous work, the Multi-Resolution Training (MRT) strategy \cite{9826014} utilizes low-resolution data to pretrain a model and subsequently transfer the pretrained model parameters to a high-resolution version of the original data, thereby reducing the overall training time.
This approach was motivated by the Multigrid method \cite{brandt1977multi}: a concept that proposes that using multiple levels of grid resolution can enhance a hierarchical problem-solving process. 
Data from scientific applications often consist of discretized physical quantities, which can be represented in different resolutions by varying the discretization granularity.
Drawing from the high-level idea of Multigrid \cite{suisalu1995adaptive, fulton1986multigrid}, at a lower resolution, a simplified approximation of the original problem could be more easily addressed.
The coarse-level solution can be incrementally refined at a finer resolution with less computational effort.
However, the MRT method is confined to pretraining only with low-resolution data and is limited to transferring the model parameters to a single type of model layer.

In this paper, we propose a novel Multi-Resolution Model Fusion (MRMF) method
capable of fusing different types of layers in a given model architecture to further accelerate the deep neural network model training.
The proposed MRMF contains a pretraining phase and a finetuning phase.
The pretraining phase has multiple model fusion stages and each stage has two models trained with data in two different resolutions.
The two trained models are then fused into a single fused model, which continues to be used as the lower-resolution model in the next stage.
The resolutions of the two sets of data gradually increase when moving from one model fusion stage to the next.
In the final finetuning stage, the fused model is trained with the original resolution data.
In addition to the model fusion contribution, we propose methods for fast low-resolution data creation and training.
We also observe the two models trained on different resolution datasets during pretraining have no dependencies on one another. 
Therefore, we investigate a concurrent model training scheme as an alternative implementation that splits the GPU resources based on estimated training time to achieve a balanced workload. 

To evaluate the performance of our proposed method, we use two real-world scientific applications, CosmoFlow \cite{mathuriya2018cosmoflow} and Neuron Inverter \cite{Ben-Shalom727974}, and analyze the performance of both applications. 
We obtain the training dataset for CosmoFlow from the MLPerf HPC training benchmark suite \cite{farrell2021mlperf, mattson2020mlperf}.
The Neuron Inverter dataset contains time-series data with billions of samples.
Our experiments were conducted on Perlmutter, a parallel computer at National Energy Research Scientific Computing (NERSC).
The results show the proposed MRMF can significantly reduce the end-to-end training time while maintaining model accuracy.
Compared with the original model training and our previous work on the multi-resolution training (MRT) method, the new method reduces end-to-end training time by up to 47\% and 23\% for CosmoFlow, respectively. 
We further observe improvements of 44\% and 24\% for Neuron Inverter, respectively.
We also present the performance evaluation of using a varying number of GPUs, from 16 to 64, verifying the proposed MRMF can maintain a near-linear scaling efficiency as the original method.
Moreover, we study the cost of generating coarse-resolution data and using different batch sizes in multi-resolution training and their impact on the training time.
Our findings show that generating coarse-resolution data incurs negligible costs for both applications.

The remainder of the paper is organized as follows:
Section~\ref{sec:background_rw} covers background information and related work.
The design and implementation of the MRMF method are detailed in Section~\ref{sec:method}.
Section~\ref{sec:eval} presents the performance results and corresponding analysis.
We conclude the paper in Section~\ref{sec:conclusion}, summarizing the main contributions of our work.

\section{Background and related works}
\label{sec:background_rw}
\subsection{Continuity in scientific applications}
In many scientific domains, datasets often consist of physical variables quantized across spatial or temporal domains, thereby exhibiting intrinsic continuity.
This inherent property allows the creation of datasets at different resolutions through selective downsampling or the aggregation of adjacent data points.
Many existing works exploit the continuity of data in the realm of scientific computing to enhance efficiency in analysis and computation. 
For instance, Suisalu applied the Multigrid method to cosmological research to solve the Poisson equation \cite{suisalu1995adaptive}. 
The Multigrid method is also used for climate modeling \cite{fulton1986multigrid}.
It is worth noting that the Multigrid method is based on a solid mathematical foundation.

We can draw parallels between the Multigrid method and Convolutional Neural Networks (CNNs) to adapt the Multigrid method for large-scale CNN training.
Previous research by He et al. \cite{he2019mgnet} makes the connection between the Multigrid method and CNN computational operations to reduce the weights and hyperparameters.
Ke et al. \cite{ke2017multigrid} proposed an extension of the Multigrid method for CNNs operating on a pyramid of spatial scales.
However, these works either present limited performance for image applications or focus on improving the final model accuracy.
In this paper, we draw inspiration from the Multigrid method by applying its principle to large-scale deep learning, aiming to enhance the efficiency of neural network training.

\subsection{Synchronous SGD with data parallelism}
In many deep learning applications, stochastic gradient descent (SGD) \cite{robbins1951stochastic} and its derivatives are commonly used to tackle optimization problems.
Mini-batch SGD \cite{kiefer1952stochastic} randomly selects a subset of the training dataset, and the model's parameters are then updated based on the gradient of the loss function computed for the mini-batch.
This process is repeated, with each mini-batch used to perform an update, until the entire dataset has been processed, which constitutes one epoch.

To enhance training efficiency, we use a synchronous parallel variant of mini-batch SGD, synchronous SGD with data parallelism, a popular parallel neural network training algorithm.
This algorithm evenly splits and distributes mini-batches to multiple workers.
Each worker independently executes computations on the assigned data to generate the gradients.
These gradients are then averaged across all participating workers through inter-process communication, ensuring a cohesive update to the model.
In this paper, we utilize synchronous SGD with data parallelism, which yields statistical efficiency, to maintain accuracy.

\subsection{CosmoFlow and Neuron Inverter benchmarks} 
We validate the MRMF method on two real-world scientific benchmarks, CosmoFlow and Neuron Inverter.
CosmoFlow applies deep learning methods in studying cosmological data.
Included in the MLPerf HPC Training benchmark\cite{mattson2020mlperf}, this project establishes a benchmark for machine learning performance evaluation on high-performance computing (HPC) systems at a large scale.
Mathuriya et al. \cite{farrell2021mlperftm,mathuriya2018cosmoflow} innovatively applied a 3D convolutional neural network model for predicting the initial conditions of the universe.
This is achieved through an analysis of 3D simulations of dark matter distribution.
CosmoFlow involves processing large multi-dimensional data, specifically analyzing 3D cubes of size $128^3$ across four redshift channels.
A major challenge of this application is the intense computational demand required for training the model on the large dataset over multiple iterations.

To comprehend neuronal activity mechanisms, simulations of single neurons can be conducted on given ion channel densities (conductances) to predict the time-series distribution of potential along various segments of the neuron (compartments) \cite{Ben-Shalom727974}.
The Neuron Inverter Machine Learning benchmark develops a deep learning tool for an inverse problem to deduce conductances in neuronal compartments from time-series data of neuronal responses \cite{balewski2022time}.
The data samples in the dataset, with dimensions $1600 \times 3$, represent the simulated response of a biological neuron across multiple compartments.
The target labels have a length of 19 and represent the conductance values used in these simulations.
The Neuron Inverter encompasses an extensive dataset with over 2 billion samples, potentially leading to an extended neural network training duration.

\section{Multi-resolution Model Fusion Method}
\label{sec:method}
The proposed Multi-Resolution Model Fusion (MRMF) method is an extension of the Multi-Resolution Training (MRT) strategy \cite{9826014}, which initially pretrains the model with data in a lower resolution and subsequently switches to train on data in the original resolution.
MRMF extends the pretraining into two phases: pretraining phase and finetuning phase.
The pretraining phase consists of multiple model fusion stages and each stage fuses two models trained with data in different resolutions.
The finetuning phase trains the model using data in the original resolution.

\subsection{Pretraining Phase}
\label{subsec:pretraining}
For later discussion, we first define a few notations.
Let $X=\{x_i: i=0,1,...N-1\}$ be the input training samples and each sample $x_i$ is an $n$-dimensional data.
Let $Y=\{y_i: i=0,1,...N-1\}$ be the labels of the corresponding training samples $X$ and each lable is $m$-dimensional.
$X$ and $Y$ together are referred to as the original dataset.
Let $\mathcal{M}$ be the deep learning model mapping $X$ to $Y$, i.e., a function $\mathcal{M}(x_i,y_i,w):\mathbb{R}^n \rightarrow \mathbb{R}^m$, where $w$ is the network weight parameters.
The original datasets $X$ and $Y$, once processed into a lower resolution, are denoted as the coarse datasets $X_c$ and $Y_c$, respectively.
Similarly, the model to be trained with the coarse datasets is referred to as {\it coarse model} denoted by $\mathcal{M}_c$.
The model to be trained with data in the higher resolution at the same fusion stage is referred to as {\it dense model} denoted by $\mathcal{M}_d$.
Our proposed model fusion method can have multiple stages and each stage fuses two models, $\mathcal{M}_c$ and $\mathcal{M}_d$.
The fused model is referred to as $\mathcal{M}_f$, which will subsequently become the coarse model in the next stage.

\textbf{Generating Data in Lower Resolutions} - 
We apply resolution reduction to both training samples $X$ and labels $Y$.
Data resolution reduction can be achieved through various well-established methods, including dimensionality reduction \cite{wold1987principal}, feature extraction \cite{zebari2020comprehensive}, compression \cite{lu2018understanding}, and downsampling.
To minimize information loss and maintain the integrity of the original model architecture, we keep the number of dimensions unchanged.
Specifically, we retain the dimension size for the channels that represent different scientific features.
For the remaining dimensions, we create a new pixel by averaging its neighbor pixels and using it to represent those neighbors.
Thus, the resolution of created coarse data depends on the number of neighbors.
This downsampling imposes a small overhead, as shown in our experiments in Section \ref{sec:discussion}.
We keep the labels $Y_c$ so that they are identical to the original data labels $Y$.

We provide an example of downsampling using the CosmoFlow dataset. In Cosmoflow, each training sample of the dataset, each representing a dark matter distribution, is stored in four cubes of size $128^3$, representing four redshifts. As the data is stored in cubes, the dimensionality we reduce is 3.
The lower resolution data is generated by averaging and replacing every $k^n$ adjacent elements in the cube, where $1 \le n \le 3$ is the number of dimensions selected to be reduced. $k\ge 2$ represents the reduced resolution.
For example, when $k=2$, the size of lower-resolution data will be $1/8$ of the original data.
As for the Neuron Inverter data set, each sample is a two-dimensional array with the first dimension denoting the time series and the second containing $3$ channels representing three compartments.
The lower-resolution data is generated by averaging every $k$ immediately adjacent elements along the second dimension, where $k\ge 2$.
 
Low-resolution data generation can be performed either on CPUs or GPUs during model training.
When running on CPUs, the reduction is executed by pooling the input data once it is loaded to CPU memory.
The generated low-resolution data is then offloaded to GPUs for training.
This approach allows caching of the reduced-resolution data in memory to be used in the successive epochs. Given there is sufficient CPU memory available, the data resolution reduction becomes a one-time operation.
For the GPU-based approach, an additional pooling layer is required at the earliest layer of the neural network architecture.
In the latter approach, the same reduction operation is repeated for every training epoch.

\begin{figure*}[t]
\centering
\includegraphics[width=1.0\textwidth]{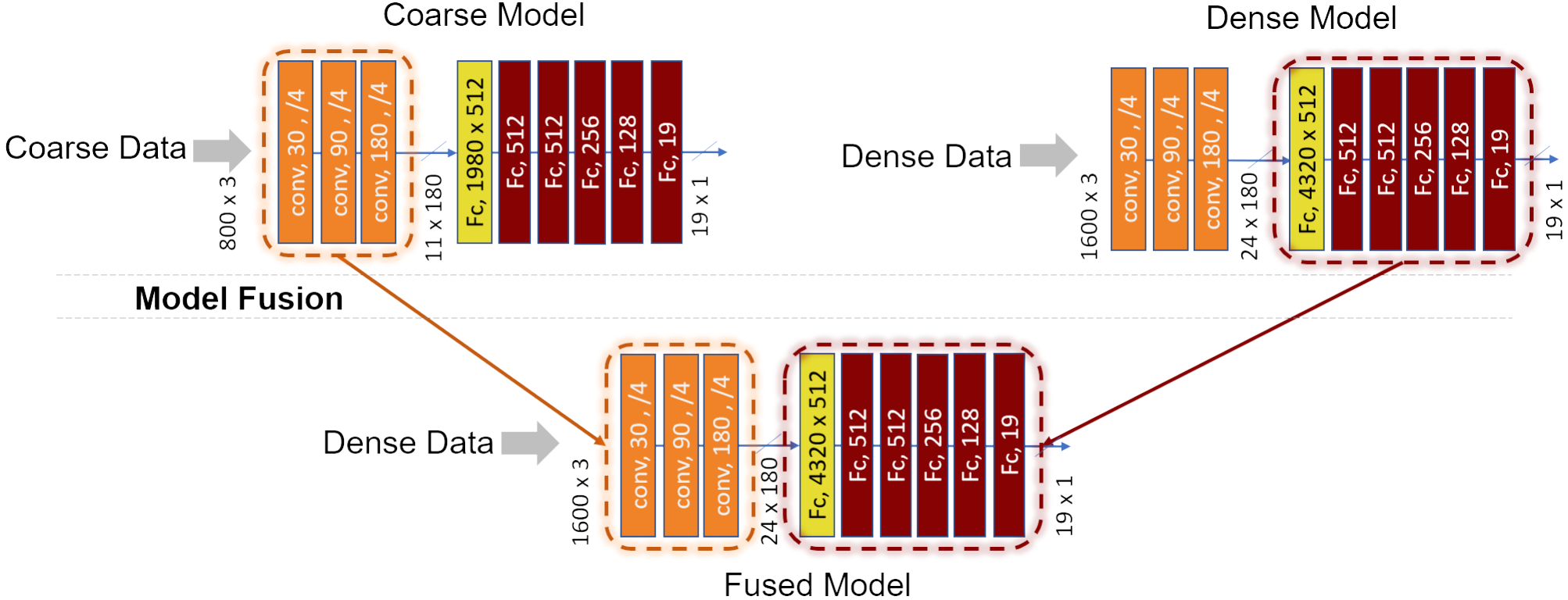}
\caption{
    Overview of Multi-resolution Model Fusion (MRMF) method. The model architecture from the Neuron Inverter benchmark is used as an example. 
    The model architectures of the coarse model $\mathcal{M}_c$ and the dense model $\mathcal{M}_d$ are on the top left and top right sides, respectively.
    The fully connected layer marked in yellow has a different weight size between the two models.
    Before model fusion, the coarse model is trained with the lower-resolution data and the dense model is trained with the higher-resolution data.
    The bottom layer group (marked in orange) from the coarse model is fused with the top layer group (marked in red) from the dense model.
    After model fusion, the fused model $\mathcal{M}_f$ is finetuned with the higher-resolution data.
}
\label{chart:model_fusion}
\end{figure*}

\textbf{Model Adjustment} - 
To use training samples in different resolutions, we adjust the model architecture.
For each fusion stage, the model trained with the data in coarse resolution is denoted by $\mathcal{M}_c$.
When creating $\mathcal{M}_c$, we keep the convolutional layers the same as the original model $\mathcal{M}$.
This allows $\mathcal{M}_c$ to process a lower-resolution input sample and generate an output size to match the dimensions of the final convolutional layer input size.
In addition, to adapt to the reduced input size, we reduce the weight size of the first fully connected layer proportionally.

As shown in the top of Figure~\ref{chart:model_fusion}, this model adjustment is exemplified with the Neuron Inverter model architecture.
Given a lower-resolution sample $x_c$ with size $800 \times 3$ ingested to the coarse model (left), the output size of the last convolutional layer is $11 \times 180 = 1980$.
Compared to the dense model (right), it is reduced from $24 \times 180 = 4320$ to $1980$.
Thus, we adjust the weight size of the first fully connected layers, marked in yellow, from $4320 \times 512$ to $1980 \times 512$.
The rest of the layers, marked in orange and red, are kept the same between the two models.
Such model adjustment is used to create the two models in each fusion stage.

\textbf{Model Fusion Stage} - 
As illustrated in Figure~\ref{chart:model_fusion}, the pretraining phase consists of one or more model fusion stages.
In each fusion stage, two models are trained with datasets in different resolutions.
The resolutions of training samples are gradually increased from one fusion stage to the next.
As depicted at the top left of Figure~\ref{chart:model_fusion}, the coarse model $\mathcal{M}_c$ is trained using the lower-resolution data (coarse data) $X_c$ and $Y_c$ for $T_c$ epochs.
Independently, the dense model $\mathcal{M}$ is trained with the higher-resolution data (dense data) for $T_d$ epochs.

For the subsequent model fusion stages, we divide the model's layers into two groups: the bottom (input side) layer group $G_b$ and the top (output side) layer group $G_t$.
For example, in the typical CNNs containing convolutional layers at the bottom and fully connected layers at the top, the convolutional layers are divided into the bottom layer group $G_b$ and the fully connected layers into the top layer group $G_t$.
Recent research on transfer learning indicates that the weights within a neural network typically converge from the bottom layer towards the top layer \cite{raghu2017svcca}.
Therefore, post this multi-resolution training, we preserve the weights of the top layer group $G_b$ from the dense model and the bottom layer group $G_t$ from the coarse model.
For other CNN model structures, the division can use a similar approach.

\subsection{Model Fusion}
At the end of each fusion stage, both trained models are combined into a fused model which becomes the coarse model in the next fusion stage.
The fusion mechanism is illustrated in Figure~\ref{chart:model_fusion}.
The top layers of the dense model $\mathcal{M}_d$ are extracted and combined with the bottom layers of the coarse model $\mathcal{M}_c$ into the fused model $\mathcal{M}_f$.
As depicted at the bottom of Figure~\ref{chart:model_fusion}, after model fusion, the higher-resolution training samples will be generated and used in the next model fusion stage.
Note that layers transferred from the coarse model are trained on lower-resolution data that contains less information than the higher-resolution data.
In the finetuning phase, the final fused model is trained with data in the original resolution till convergence.

This model fusion strategy provides a better foundation in the early stage of model training and allows utilization of the knowledge gained from various data resolutions.
There is empirical evidence on image datasets suggesting that the features extracted in the first few layers resemble each other across various datasets \cite{zeiler2014visualizing}. 
Furthermore, it is known that the model accuracy can be improved by transferring features even from distant tasks \cite{yosinski2014transferable}.
Therefore, we expect the fused model to achieve similar accuracy as the original-resolution model but require fewer training epochs in the finetuning phase.
Together with the fusion phase, which runs faster as lower-resolution data demands less computational cost, MRMF can effectively reduce the end-to-end training time without sacrificing the end model accuracy.

\textbf{Training Stop Condition}
\label{subsec:switching} - 
One of the important mechanisms in the model fusion is to determine the duration of the model training.
Extending the training duration on lower-resolution data could potentially enhance knowledge acquisition but at the expense of increased computational costs.
However, without sufficient training, knowledge transferred through the model fusion may not effectively reduce the subsequent training time.
We implement a stop condition based on the training loss so that it makes a efficient trade-off between accuracy and efficiency.
By monitoring the variance in training losses of adjacent epochs, we use a threshold $\epsilon$ and a patience $T_p$ to decide whether or not further training would lead to a fast loss reduction.
If the training loss reduction is smaller than the threshold $\epsilon$ for a consecutive of $T_p$ epochs, we stop the training and start the model fusion.
This threshold is set based on the loss decline in the early training epochs, which can be tuned for different applications and data resolutions.

\begin{figure*}[t]
\centering
\includegraphics[width=1.0\textwidth]{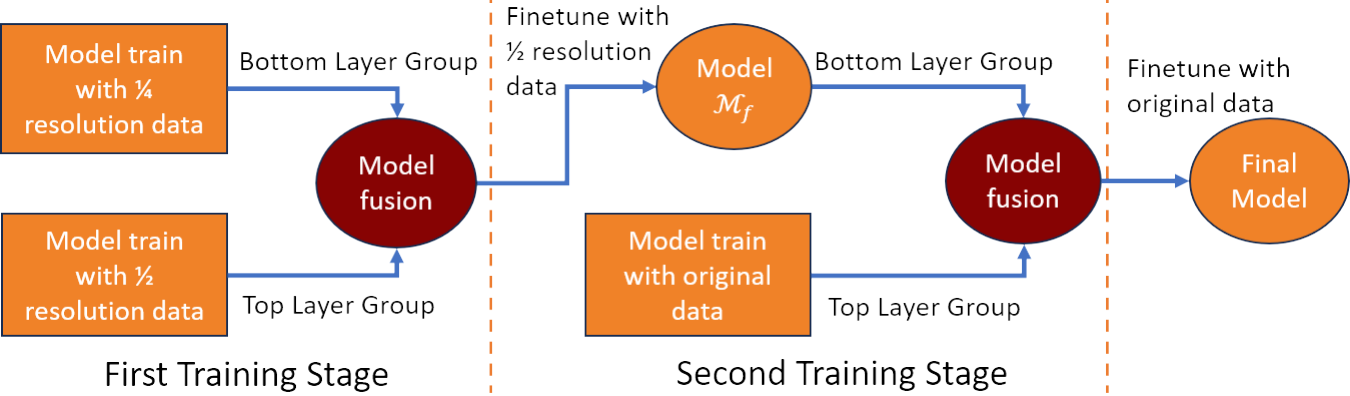}
\caption{
    The pipeline of extending the pretraining phase into multiple stages to use progressively increased two different resolution data sets to conduct two-model training in each stage. 
    The top layers of the higher-resolution model and the bottom layers of the lower-resolution model are then combined into the fused model for the next stage.
}
\label{chart:multi-fusion}
\end{figure*}

\subsection{Pipeline of Multiple Model Fusions}
\label{subsec:multi-fusion}
Note that the inherent continuity in scientific data allows the discretization of physical quantities at different resolutions.
We design a progressive multi-fusion training method to extend the one-fusion MRMF shown in Figure~\ref{chart:model_fusion}.
As shown in Figure~\ref{chart:multi-fusion}, we exemplify this method with three resolutions of the original data with size ratios of $\frac{1}{4}$, $\frac{1}{2}$, and $1$. 
Based on the MRMF described in the previous section, in the first training stage, we treat data of $\frac{1}{4}$ and $\frac{1}{2}$ resolutions of the original data as the lower-resolution and the higher-resolution data and train them on each of the models.
Once the stop conditions of the models are satisfied, we conduct model fusion on these two models by merging them into the fused model $\mathcal{M}_f$, which is able to be trained with $\frac{1}{2}$ resolution data. 
Similarly, in the second fusion stage, the data in the $\frac{1}{2}$ and original resolutions are considered the lower-resolution and the higher-resolution data, respectively.
The fused model $\mathcal{M}_f$ in the previous stage is used as the coarse model and trained with the lower-resolution data, while the dense model is trained with the original-resolution data.
After both models stop their training, the second model fusion is conducted to fuse them into the final fused model for finetuning with the original data.
When generalizing to different applications, the number of training stages and data resolutions used in each stage can be adjusted based on the original data sample size.

\subsection{Parallel and Concurrent Multi-Resolution Training}
\label{subsec:concurrent}
In MRMF, we implement synchronous SGD with data parallelism to parallelize the neural network training.
Data samples are distributed evenly across all workers, enabling simultaneous training on individual local models.
At the end of each iteration, an inter-process communication is conducted to average the gradients across all workers.
This aggregation process yields globally averaged gradients, which are used to update each local model for the next epoch to ensure model synchronization.
The same parallelization is used in training individual models.

\textbf{Concurrent Model Training} - 
In each fusion stage, the two models can be trained independently from each other.
This is based on the fact that training the dense model $\mathcal{M}$ is using data separately from training the coarse model $\mathcal{M}_c$.
In each fusion stage, training of the dense and coarse models can be done one after another using all allocated computing resources, or concurrently by dividing the resources between the two trainings.
For the latter, we can divide all available computational resources into two groups: group dense $G_d$ and group coarse $G_c$. 
In the training stage before model fusion, the workers in group $G_d$ are used to train the dense model, while the rest of the workers from group $G_c$ are used to train the coarse model concurrently.

To obtain a high efficiency of concurrent training, our goal is to divide the computational resources so that it can yield a balanced workload among all workers.
Assuming the training times for the dense and coarse models are $t_d$ and $t_c$, respectively, we use the allocation ratios of $\frac{t_d}{t_d + t_c}$ for the dense model and $\frac{t_c}{t_d + t_c}$ for the coarse model.
Given a fixed number of computing resources, we expect this workload division mechanism to achieve a similar training time as the sequential training mechanism.

\section{Evaluation}
\label{sec:eval}
We evaluated MRMF using two real-world scientific CNN applications: CosmoFlow and Neuron Inverter. 
MRMF is implemented in Python using Tensorflow \cite{abadi2016tensorflow} and PyTorch \cite{paszke2019pytorch} Deep Learning Framework.
All experiments were performed on Perlmutter, an HPE Cray EX supercomputer at the National Energy Research Scientific Computing Center \cite{nersc_pm,nersc_pmgpu}.
Each compute node has a 64-core single AMD EPYC 7763 (Milan) CPU and 4 NVIDIA A100 (Ampere) GPUs.
There are 256 GB of DDR4 DRAM per node.
The system is equipped with an all-flash file system and high-speed interconnections with a 3-hop dragonfly topology.

\textbf{Software} - 
CosmoFlow is developed using TensorFlow 2.6.0 and Horovod 0.24.3. 
Neuron Inverter is developed using PyTorch 2.0.1 with DistributedDataParallel (DDP) module.
Both applications have been parallelized using NCCL 2.15.5 and cudnn 8.3.2.
In our experiments, we evaluated their performance using multiple GPUs, ranging from 16 to 64, with one MPI rank per GPU allocated.

\subsection{Performance Results of CosmoFlow}
We use a modified version of CosmoFlow that is based on Livermore Big Artificial Neural Network (LBANN) \cite{van2015lbann}, which contains seven 3-D convolutional layers and three fully connected layers.
The input data set contains 9998 samples, totaling 157 GB.
Each data sample of CosmoFlow is of size $128^3 \times 4$, which is binned from the simulated raw data of size $512^3 \times 4$. 
This original data serves as the same source for building the CosmoFlow data in the MLPerf HPC benchmark \cite{farrell2021mlperftm}.
The first three dimensions of a sample represent 3D matter distribution, while the fourth dimension represents four channels of redshifts of the evolved universe.
All data samples are stored in 80 HDF5 files, which are randomly divided into 80\% for training, 10\% for validation, and 10\% for testing.
The training sample batch size is set to 8 per GPU.

We refer to the \textbf{baseline} as training the unmodified model on the original data in our result analysis. 
Given a set of optimized hyperparameters, the baseline model achieves a maximum Mean Squared Error (MSE) value of 0.0025. 
Based on the strategy for generating coarse data described in Section \ref{subsec:pretraining}, due to the semantic significance of each channel in the fourth dimension, the data reduction is only applied to the three spatial dimensions.
We generated the lower-resolution data samples (coarse data) by averaging every $2^3$ adjacent elements from the first three dimensions.
For comparison, we report the end-to-end training time of CosmoFlow using our Multi-Resolution Model Fusion (MRMF) method with one model fusion to compare with the baseline method as well as the Multi-Resolution Training (MRT) method proposed in our previous work \cite{9826014}.
All results are averaged over three runs using random seeds in each training method.

\subsubsection{End-to-end Training}

\begin{table}[t] 
\caption{The validation loss, number of training epochs before fusion, number of training epochs after fusion, end-to-end training time (on Perlmutter) for CosmoFlow with 32 GPUs. The timings are all in seconds.}
\begin{center}
\begin{tabular}{|c||c|c|c|}
\hline
\textbf{Training method} & \textbf{Baseline} & \textbf{MRT} & \bfseries\makecell[c]{MRMF \\(One Fusion)}\\
\hline
\hline
\bfseries\makecell[c]{Validation \\loss (MSE)} & 0.0025 & 0.0025 & \textbf{0.0024}\\
\hline
\hline
\bfseries\makecell[c]{Number of\\ epochs before fusion} & - & 55 (coarse) & \makecell[c]{52 (coarse) \\+ 7 (dense)}\\
\hline
\bfseries\makecell[c]{Number of \\ epochs after fusion} & 88 & 54 (dense) & 38 (dense)\\
\hline
\hline
\bfseries\makecell[c]{Total time \\ (Perlmutter)} & 2181.61 & 1498.82 & \textbf{1258.43}\\
\hline
\end{tabular}
\label{table:cos-time-32}
\end{center}
\end{table}

\begin{table}[t] 
\caption{The validation loss, number of training epochs before fusion, number of training epochs after fusion, end-to-end training time (on Perlmutter) for CosmoFlow with 64 GPUs. The timings are all in seconds.}
\begin{center}
\begin{tabular}{|c||c|c|c|}
\hline
\textbf{Training method} & \textbf{Baseline} & \textbf{MRT} & \bfseries\makecell[c]{MRMF \\(One Fusion)}\\
\hline
\hline
\bfseries\makecell[c]{Validation \\loss (MSE)} & 0.0025 & 0.0025 & \textbf{0.0024}\\
\hline
\hline
\bfseries\makecell[c]{Number of\\ epochs before fusion} & - & 55 (coarse) & \makecell[c]{52 (coarse) \\+ 8 (dense)}\\
\hline
\bfseries\makecell[c]{Number of \\ epochs after fusion} & 88 & 54 (dense) & 37 (dense)\\
\hline
\hline
\bfseries\makecell[c]{Total time \\ (Perlmutter)} & 1354.04 & 763.31 & \textbf{681.17}\\
\hline
\end{tabular}
\label{table:cos-time-64}
\end{center}
\end{table}

Table~\ref{table:cos-time-32} shows the results of the baseline, MRT, and MRMF running on 32 GPUs.
The baseline method takes 88 epochs to converge to the final model accuracy.
In the previously proposed MRT approach, the model was trained on the lower resolution (coarse) data for 55 epochs and subsequently trained on the original (dense) data for 54 epochs.
In contrast,
the model was pretrained for 52 epochs with the coarse data, 7 epochs with the dense data, and finally 38 epochs with the dense data after model fusion.
All three training methods used the same model accuracy convergence condition, a validation loss of 0.0025.
These experiments indicate that the parameters learned before fusion help reduce the number of training epochs required for later training with dense data.
Among the three training methods, MRMF clearly achieves the shortest training time, showing improvements of 42.3\% and 16.0\% over the baseline and MRT, respectively.

To evaluate the impact of model fusion, we further compare the timing results between MRMF and MRT.
Based on the switching mechanism described in Section \ref{subsec:switching}, we set the threshold and patience to $0.002$ and $5$ as the stop condition of the lower resolution model.
For the dense model, a stricter condition with threshold $0.005$ and patience $3$ is used since a higher epoch time is expected.
For MRT, $97.46$ seconds are used to pretrain with the lower-resolution data, while $1401.34$ seconds are for finetuning with the dense data.
For MRMF, pretraining with the lower-resolution data costs $91.52$ seconds, and pretraining with the dense data costs $197.61$ seconds.
Applying model fusion to the pretrained model leads to a shorter dense data finetuning time of $969.30$ seconds.
This improvement demonstrates that the proposed model fusion can provide a better foundation for finetuning compared to MRT.

\begin{table}[t] 
\caption{The training time breakdown of one fusion and two fusion methods for CosmoFlow (on Perlmutter) with 32 GPUs. The timings are all in seconds.}
\begin{center}
\begin{tabular}{|c||c|c|}
\hline
\textbf{ } & \bfseries\makecell[c]{One Fusion} & \bfseries\makecell[c]{Two Fusions} \\
\hline
\hline
\textbf{\makecell{First\\ training stage}} & 289.11 & 135.49 \\
\hline
\hline
\textbf{\makecell{Second\\ training stage}} & - & 328.52 \\
\hline
\hline
\textbf{Finetuning stage} & 969.30 & 683.55 \\
\hline
\hline
\hline
\textbf{Total time} & 1258.43 & \textbf{1147.59} \\
\hline
\hline
\textbf{Validation loss (MSE)} & 0.0024 & \textbf{0.0024} \\
\hline
\end{tabular}
\label{table:cos-time-multi}
\end{center}
\end{table}

\subsubsection{Multi-fusion}
As illustrated in Figure~\ref{chart:multi-fusion}, MRMF can be extended to multiple fusions, each with two sets of progressively increased resolution of data.
We evaluate this multi-fusion method on CosmoFlow with three resolutions of data in three stages: the first training stage, the second training stage, and the final model finetuning stage.
Considering the coarse data used in one-fusion training is $\frac{1}{8}$ of the original data with a small average epoch time. 
In multi-fusion, we use $\frac{1}{8}$, $\frac{1}{2}$, and original resolutions of data.
Note that due to the first three dimensions of CosmoFlow data representing the three dimensions in matter distribution, $\frac{1}{2}$ resolution allows pooling on one of them. 
Thus, we present the average results of pooling across each of the three dimensions, represented as $\frac{1}{2}$ resolution data.
In the first training stage, we use $\frac{1}{8}$ and $\frac{1}{2}$ resolutions of the original data as the coarse and the dense data.
After the first model fusion, the fused model continues to be used as the coarse model.
In the second training stage, $\frac{1}{2}$ and the original resolutions of data are used as the coarse and the dense data.
In the final stage, the final fused model is finetuned with the original data.

Table~\ref{table:cos-time-multi} shows the training time breakdown of the one-fusion and the two-fusion MRMF. 
Given the expectation of a larger loss reduction in the early training stage, we set the switching threshold of $\frac{1}{2}$ resolution model to $0.01$ and $0.001$ in the first and the second training stages, respectively.
We can see that in two-fusion training, the training time clocks in at $135.49$ seconds, which is shorter compared to $289.11$ seconds in one-fusion training due to the smaller dense data size.
Then, the training time gradually increases in each stage as a pair of higher resolutions of data is used.
By using two-fusion training, the finetuning time in the final stage is reduced from $969.30$ seconds to $683.55$ seconds, resulting in the end-to-end training time being reduced by $8.81 \%$ compared to one-fusion training.
These comparison results affirm the effectiveness of multi-fusion training in further reducing the training time.

\subsubsection{Scaling Performance}
To evaluate the impact of MRMF on scalability, we measure the strong scaling results of CosmoFlow and then analyze the timing breakdown.
When increasing the number of GPUs used, we keep the global batch size unchanged as $256$.
Note that the model does not fit into the memory when using a number of GPUs smaller than 32, the scaling performance is presented from 32 GPUs.

Table~\ref{table:cos-time-64} shows the end-to-end training time of CosmoFlow using 64 GPUs.
With the experiment setup employed in Table~\ref{table:cos-time-32}, we compare the results of using MRMF with the baseline and MRT.
Due to the same hyperparameter settings, all three methods achieve the model accuracy converge condition, 0.0025 validation loss, with similar numbers of training epochs before and after model fusion as the results on 32 GPUs.
As shown from the baseline, end-to-end time (1354.02 seconds) is reduced to 62.06\% when scaling from 32 GPUs to 64 GPUs, MRMF further reduces the time to 681.17 seconds.
The total time reduction of MRMF on 64 GPUs is 49.69\% compared to the baseline and 10.76\% compared to MRT.

\begin{figure}[t]
\centering
\includegraphics[width=.65\textwidth]{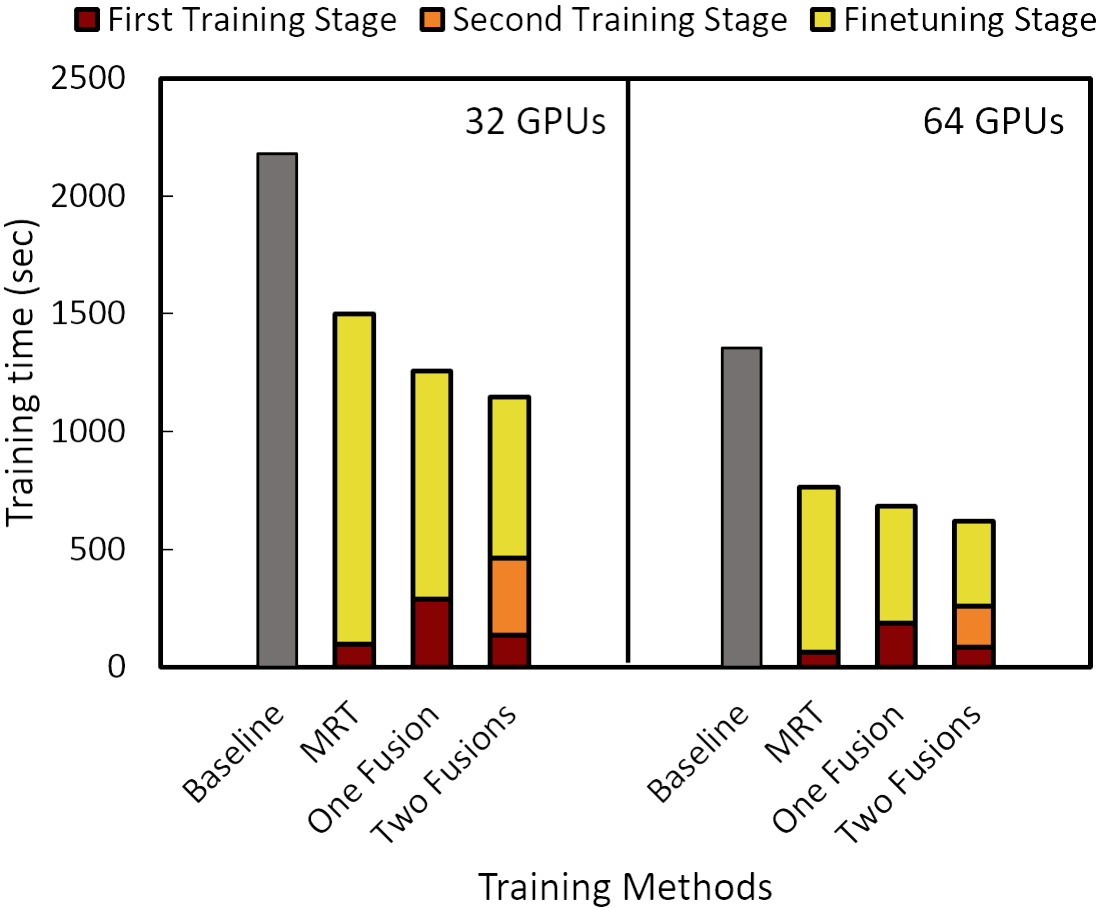}

\caption{
    Training timing breakdown for CosmoFlow comparing the baseline (the original model), the MRT method, and the proposed MRMF method using one fusion and two fusions. 
    The experiments are conducted on both 32 and 64 GPUs on Perlmutter.
}
\label{chart:cos_time}
\end{figure}

\begin{table}[t]
\caption{The average epoch timing breakdown for CosmoFlow on Perlmutter. The timings are all in seconds.}
\begin{center}
\begin{tabular}{|c|c||c|c|c|c|}
\hline
\bfseries\makecell[c]{Number \\of GPUs} & \textbf{Dataset} & \bfseries\makecell[c]{I/O \\time} & \bfseries\makecell[c]{Comp \\time} & \bfseries\makecell[c]{Comm \\time}  & \bfseries\makecell[c]{Average \\epoch time} \\
\hline
\hline
\multirow{2}{*}{32} & \textbf{Coarse} & 0.25 & 1.08 & 0.43  & 1.76 \\
\cline{2-6}
& \textbf{Dense} & 16.77 & 8.95 & 0.99 & 26.71 \\
\hline
\hline
\multirow{2}{*}{64} & \textbf{Coarse} & 0.15 & 0.54 & 0.13 & 0.82 \\
\cline{2-6}
& \textbf{Dense} & 7.01 & 4.46 & 1.77 & 13.24 \\
\hline
\end{tabular}
\label{table:cos-time-breakdown}
\end{center}
\end{table}

Figure~\ref{chart:cos_time} shows the training time of CosmoFlow using 32 and 64 GPUs, in particular the differences among four training methods: the baseline, MRT, one-fusion MRMF, and two-fusion MRMF.
The training time in the first, second, and finetuning stages is marked in red, orange, and yellow respectively.
The chart shows that MRMF with two model fusions achieves the shortest training time among the four methods.
As the training progresses from the first stage to the final finetuning stage, the training time gradually increases with the increasing resolution of the training data.
The pretraining phase on lower-resolution data, denoted in red and orange, effectively reduces the time required for finetuning, marked in yellow, required on the original data to maintain the model accuracy.
Furthermore, when the GPU count is increased from 32 to 64, MRMF continues to outperform the baseline and MRT while almost scaling linearly.

We analyze the timing breakdown using 32 and 64 GPUs on both coarse and dense data in one-fusion MRMF, which is shown in Table~\ref{table:cos-time-breakdown}.
In one-fusion MRMF, the coarse data is at $\frac{1}{8}$ of the original resolution, while the dense data is the original data.
First, Table~\ref{table:cos-time-breakdown} shows that the I/O time dominates the average epoch time on the original (dense) data.
This time is nearly reduced by one-half when using 64 GPUs because fewer samples are assigned to each process.
The I/O time is nearly negligible when training on the coarse data due to the much smaller sample size.
Second, as the sample size in the coarse data is $\frac{1}{8}$ of the dense data, computation time is significantly reduced. 
Moreover, the communication time takes a shorter time due to the smaller weight size in the fully connected layers.
Thus, the training on the lower-resolution data has a minimal average epoch time, leading to a small cost of pretraining before the model fusion.
As MRMF effectively reduces the number of training epochs required for model convergence in the finetuning stage and incurs a negligible pretraining overhead, MRMF shows potential for compatibility with many other scalable approaches.

We further evaluate the concurrent model training described in Section \ref{subsec:concurrent} using one-fusion MRMF.
The proposed concurrent model training method can be used as an alternative to sequential pretraining.%
When using 32 GPUs, we divide the GPUs into two groups: 20 GPUs in group dense $G_d$ and 12 GPUs in group coarse $G_c$ to pretrain both dense and coarse models concurrently.
Some samples were excluded from dense pretraining due to indivisibility by the number of GPUs.
The concurrent model training achieves an end-to-end time of $1249.47$ seconds, closely matching the sequential training time of $1258.43$ seconds.
Scaling up to 64 GPUs, 44 GPUs are distributed to group dense $G_d$ and 20 GPUs are distributed to group coarse $G_c$.
To our expectation, concurrent model training reaches the same target validation loss of $0.0025$ in $678.05$ seconds, similar to the sequential training result of $681.17$ seconds. 
These marginal timing improvements are likely due to sample exclusions and the reduced GPU count per task.
These results confirm the viability of concurrent training as an effective alternative implementation.

\subsection{Performance Results of Neuron Inverter}
The data set used in Neuron Inverter is generated from a simulated biological neuron \cite{balewski2022time}. 
The model consists of three 1-D convolutional layers, a batch normalization layer, and five fully connected layers.
The input data set contains 26,251,750 samples, dividing into 21,001,400 for training, 2,625,175 for validation, and 2,625,175 for testing.
The total size is 472 GB.
Each sample is of size $1600 \times 3$.
The first dimension represents time-series neuronal responses, while the three channels represent the three compartments of a biological neuron.
All data samples are stored in an HDF5 file.
Samples are divided into 80\% for training, 10\% for validation, and 10\% for testing.
The training sample batch size is set to 512 per GPU.
Similar to CosmoFlow, we refer to the original Neuron-Inverter dataset trained on the original model as the \textbf{baseline}.
Based on the coarse data generating mechanism proposed in Section \ref{subsec:pretraining}, we average every adjacent $2$ values in the first dimension to generate the coarse data, each with a size of $800 \times 3$.

Following the settings from the Neuron-Inverter benchmark \cite{balewski2022time}, we trained the model using Adam optimizer \cite{kingma2014adam}.
During training initialization, the Neuron Inverter data is evenly divided and assigned to each process, and the data is preloaded into local memory to minimize the I/O overhead during training.
Due to GPU node memory limitation, we evaluate the performance on Perlmutter using 16 and 32 GPUs without using concurrent model training.

\subsubsection{End-to-end Training}

\begin{table}[t] 
\caption{The validation loss, number of training epochs before fusion, number of training epochs after fusion, end-to-end training time (on Perlmutter) for Neuron Inverter with 16 GPUs. The timings are all in seconds. }
\begin{center}
\begin{tabular}{|c||c|c|c|}
\hline
\textbf{Training method} & \textbf{Baseline} & \textbf{MRT} & \bfseries\makecell[c]{MRMF \\(One Fusion)}\\
\hline
\hline
\bfseries\makecell[c]{Validation \\loss (MSE)} & 0.0400 & 0.0399 & \textbf{0.0398}\\
\hline
\hline
\bfseries\makecell[c]{Number of\\ epochs before fusion} & - & 35 (coarse) & \makecell[c]{35 (coarse) \\+ 4 (dense)}\\
\hline
\bfseries\makecell[c]{Number of \\ epochs after fusion} & 84 & 41 (dense) & 32 (dense)\\
\hline
\hline
\bfseries\makecell[c]{Total time \\ (Perlmutter)} & 3840.60 & 2802.61 & \textbf{2528.51}\\
\hline
\end{tabular}
\label{table:ni-time-16}
\end{center}
\end{table}

\begin{table}[t]
\caption{The validation loss, number of training epochs before fusion, number of training epochs after fusion, end-to-end training time (on Perlmutter) for Neuron Inverter with 32 GPUs. The timings are all in seconds.}
\begin{center}
\begin{tabular}{|c||c|c|c|}
\hline
\textbf{Training method} & \textbf{Baseline} & \textbf{MRT} & \bfseries\makecell[c]{MRMF \\(One Fusion)}\\
\hline
\hline
\bfseries\makecell[c]{Validation \\loss (MSE)} & 0.0400 & 0.0399 & \textbf{0.0399}\\
\hline
\hline
\bfseries\makecell[c]{Number of\\ epochs before fusion} & - & 34 (coarse) & \makecell[c]{34 (coarse) \\+ 4 (dense)}\\
\hline
\bfseries\makecell[c]{Number of \\ epochs after fusion} & 83 & 41 (dense) & 32 (dense)\\
\hline
\hline
\bfseries\makecell[c]{Total time \\ (Perlmutter)} & 2045.90 & 1553.31 & \textbf{1418.01}\\
\hline
\end{tabular}
\label{table:ni-time-32}
\end{center}
\end{table}

Table~\ref{table:ni-time-16} presents the results of the baseline, MRT, and MRMF with one model fusion on 16 GPUs.
For the baseline, the number of epochs required until convergence is 84.
For MRT, the number of epochs required during the pretraining with the lower resolution data is 35, which is followed by 41 epochs of training with the original (dense) data.
For MRMF, the pretraining stage consists of 35-epoch training with the lower resolution data and 4-epoch training with the dense data, which is followed by 32-epoch training with dense data after the model fusion.
All three training methods use the same model accuracy converge condition, 0.0400 validation loss.
In comparison to the baseline, MRMF shows an improvement of $34.16\%$ of the training time.
Compared with MRT, both methods transfer weights of convolutional layers from the model pretrained on the lower-resolution data, while MRMF further utilizes weights from pretraining on the dense data.
The timing reduction indicates that knowledge transferred from the dense model helps to reduce the number of training epochs after model fusion and reduces the end-to-end training time by $9.78\%$ on 16 GPUs.
Similarly, in Table~\ref{table:ni-time-32}, MRMF clearly achieves the shortest training time among the three training methods when running on 32 GPUs.
The combination of these results demonstrate the effectiveness of MRMF in time-series problems for reducing the time to achieve the same validation accuracy.

\subsubsection{Multi-fusion}

\begin{table}[t] 
\caption{The training time breakdown of one fusion and two fusion methods for Neuron Inverter (on Perlmutter) on 16 GPUs. The timings are all in seconds.}
\begin{center}
\begin{tabular}{|c||c|c|}
\hline
\textbf{ } & \bfseries\makecell[c]{One Fusion} & \bfseries\makecell[c]{Two Fusions} \\
\hline
\hline
\textbf{\makecell{First\\ training stage}} & 709.20 & 166.35 \\
\hline
\hline
\textbf{\makecell{Second\\ training stage}} & - & 525.60 \\
\hline
\hline
\textbf{Finetuning stage} & 1541.00 & 1421.50 \\
\hline
\hline
\hline
\textbf{Total time} & 2250.21 & \textbf{2113.45} \\
\hline
\hline
\textbf{Validation loss (MSE)} & 0.040 & \textbf{0.039} \\
\hline
\end{tabular}
\label{table:ni-time-multi}
\end{center}
\end{table}

Based on the multi-fusion training described in Section \ref{subsec:multi-fusion}, we test multi-fusion on the Neuron Inverter benchmark with three data resolutions: $\frac{1}{4}$, $\frac{1}{2}$, and the original resolution.
In the first training stage, we train the model with samples of $\frac{1}{4}$ resolution and $\frac{1}{2}$ resolution as the coarse and the dense model.
After training, the first model fusion is conducted on the two trained models.
Then, in the second training stage, this fused model is used as the coarse model to be trained with samples of $\frac{1}{2}$ resolution, while the dense model is trained with the original-resolution data.
After the second model fusion, this final fused model is finetuned with the original dataset.

Table~\ref{table:ni-time-multi} shows the training time comparison between the one-fusion and the two-fusion training on 16 GPUs. 
The local batch size is set to 2048, 2048, and 512 for samples of $\frac{1}{4}$, $\frac{1}{2}$, and original resolution.
We set $3$ to patience and $0.0002$ to the threshold as the stopping condition of pretraining on the original and $\frac{1}{2}$ resolution of data.
As a larger loss reduction is expected in the early stages of training, we increase the switching threshold to $0.002$ as the stop condition for the first training stage.
In one-fusion training, the pretraining time is $709.20$ seconds, with both $\frac{1}{2}$ and the original resolution of data.
In two-fusion training, the pretraining stage consists of $166.35$ seconds on pairs of lower-resolution data and $525.60$ seconds on pairs of higher-resolution data.
In comparison to only using the two-resolution data, pretraining with data of $\frac{1}{4}$ resolution helps to reduce the pretraining time required on data of $\frac{1}{2}$ size in the second training stage.
Furthemore, the training time in the final model finetuning stage is reduced from $1541.00$ seconds to $1421.50$ seconds, which shows that using data with more resolutions in the early stage provides a better foundation for susequent training.

\subsubsection{Scaling Performance}

\begin{figure}[t]
\centering
\includegraphics[width=.65\textwidth]{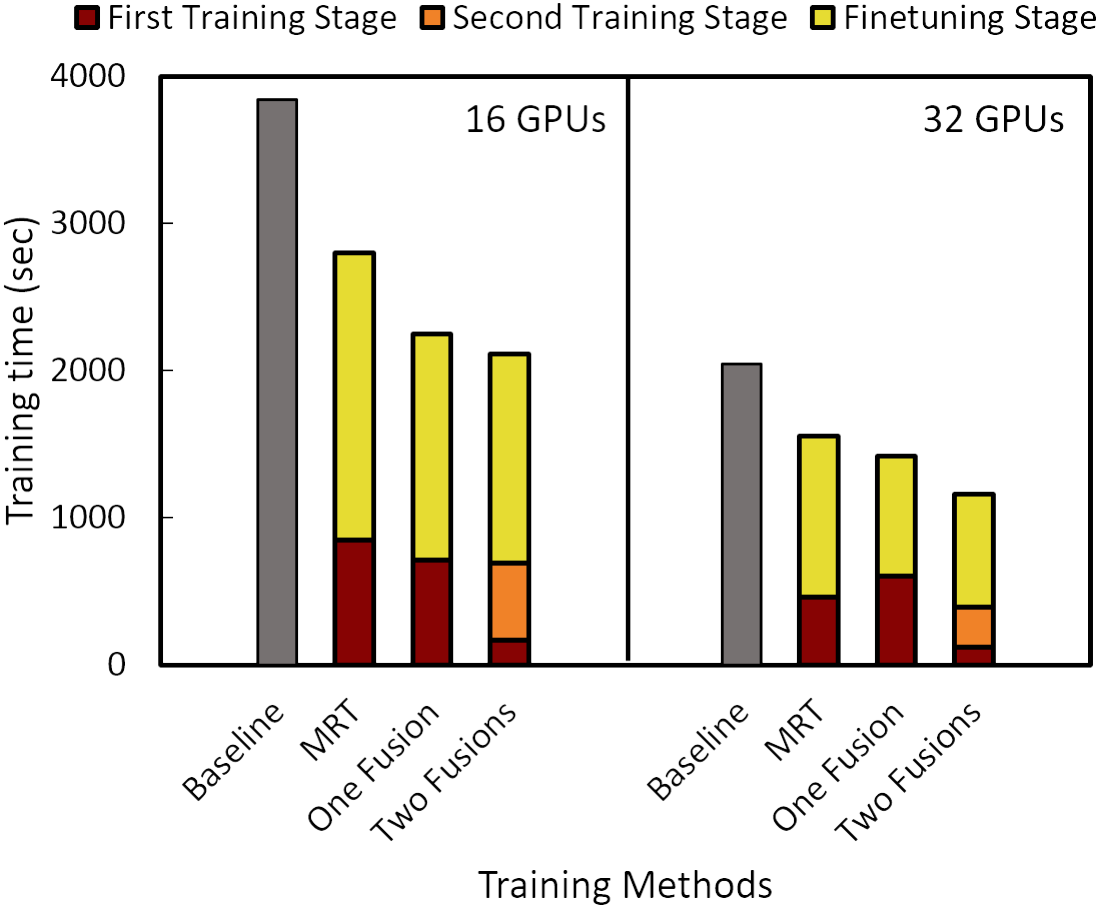}

\caption{
    Comparison of the training timing breakdown for Neuron Inverter between baseline, the MRT method, and the proposed MRMF method with one fusion and two fusions on Perlmutter. 
    The experiments are conducted on both 16 and 32 GPUs.
}
\label{chart:ni_time}
\end{figure}

Figure~\ref{chart:ni_time} presents the scaling performance of the Neuron Inverter benchmark on Perlmutter GPU nodes.
We compare the results of using 16 and 32 GPUs among four training methods: the baseline, MRT, one-fusion MRMF, and two-fusion MRMF.
Although training in pretraining phase, marked in red and orange, incurs an additional overhead, the training before model fusion potentially reduces the time until convergence following the model fusion phase for both GPU settings.
By using two-fusion training, MRMF shows improvements of 44.97\% and 24.59\% over the baseline and MRT, respectively.
When scaled to 32 GPUs, the chart shows that MRMF continues to outperform the baseline and MRT.

\begin{table}[t]
\caption{The average epoch timing breakdown for Neuron Inverter on Perlmutter. The timings are all in seconds.}
\begin{center}
\begin{tabular}{|c|c||c|c|c|}
\hline
\bfseries\makecell[c]{Number \\of GPUs} & \textbf{Dataset} & \bfseries\makecell[c]{Comp \\time} & \bfseries\makecell[c]{Comm \\time} & \bfseries\makecell[c]{Average \\epoch time} \\
\hline
\hline
\multirow{2}{*}{16} & \textbf{Coarse} & 10.83 & 10.83 & 20.03 \\
\cline{2-5}
& \textbf{Dense} & 19.89 & 21.43 & 41.32 \\
\hline
\hline
\multirow{2}{*}{32} & \textbf{Coarse} & 5.63 & 5.43  & 11.06 \\
\cline{2-5}
& \textbf{Dense} & 10.11 & 14.07 & 24.18 \\
\hline
\end{tabular}
\label{table:ni-time-bd}
\end{center}
\end{table}

Table~\ref{table:ni-time-bd} presents the timing breakdown of training on the coarse and dense Neuron Inverter datasets for a more in-depth analysis of the MRMF method.
The coarse and dense datasets are data of $\frac{1}{2}$ and original resolutions used in one-fusion MRMF training.
Initially, we observe an I/O overhead in the first epoch when data samples are preloaded to local memory by each I/O process and shuffled locally.
As the size of the coarse dataset is half the size of the dense dataset, the corresponding I/O time decreases proportionally to nearly half of the previous I/O time.
Due to in-memory data loading, the I/O time becomes negligible after the first epoch; thus, is not included in the table. 
Second, we observe increasing the number of GPU devices reduces the average communication time per epoch for both data resolutions.
The reason behind the reduced communication time is because the local batch size remains constant causing the global batch size to double. This causes the number of iterations per epoch to reduce proportionally, therefore lowering the frequency of gradient averaging per epoch.
Finally, training on the coarse dataset reduces the computational time due to the reduced input data size.

\subsection{Discussion}
\label{sec:discussion}
\subsubsection{Impact of Lower Resolution Data Generation}
In this subsection, we present an analysis of the impact of using the mechanism introduced in Section \ref{subsec:pretraining} to generate lower-resolution data.
To investigate the impact during the model pretraining phase, we compare the average epoch time of training on original data with the time of conducting data resolution reduction on original data.
As introduced in Section \ref{subsec:pretraining}, low-resolution data can be generated via two methods: CPU-based processing or adding an extra pooling layer to conduct GPU-based processing. 

\begin{table}[t] 
\caption{The average epoch time (original model), time of pooling on CPUs and GPUs (on Perlmutter) for CosmoFlow and Neuron Inverter. The timings are all in seconds. }
\begin{center}
\begin{tabular}{|c||c|c|}
\hline
\textbf{ } & \bfseries\makecell[c]{CosmoFlow} & \bfseries\makecell[c]{Neuron Inverter} \\
\hline
\hline
\textbf{Average epoch time} & 26.71 & 41.32 \\
\hline
\hline
\textbf{Pooling on CPUs} & 6.46 & 1.87 \\
\hline
\hline
\textbf{Pooling on GPUs} & 4.02 & 1.80 \\
\hline
\end{tabular}
\label{table:pooling}
\end{center}
\end{table}

Table~\ref{table:pooling} shows the average epoch time of the original model and data resolution reduction time on CPUs and GPUs for both two applications.
For CosmoFlow, the pooling is conducted on the original data, aiming to generate data of $\frac{1}{8}$ resolution using 32 GPUs.
For Neuron Inverter, the original data is pooled to create $\frac{1}{2}$ resolution data with 16 GPUs.
We observe that conducting pooling on CosmoFlow data through CPUs and GPUs costs $6.46$ seconds and $4.02$ seconds respectively, which is much less than the average epoch time.
Similarly, for the Neuron Inverter, reducing the data resolution allows for marginal reductions to the epoch time, $1.87$ seconds and $1.80$ seconds respectively, due to the low data dimensions.
Moreover, using a pooling layer to reduce data resolution on GPUs is faster than pooling with CPUs in both applications as previously hypothesized.
Note that the generated low-resolution data can be kept in CPU memory when there is enough CPU memory space.
Thus, the cost of data resolution reduction on CPUs could be further reduced to a one-time cost by generating and storing the low-resolution data at the start of training.

\subsubsection{Impact of Local Batch Size in Pretraining}
\begin{figure}[t]
\centering
\includegraphics[width=.65\textwidth]{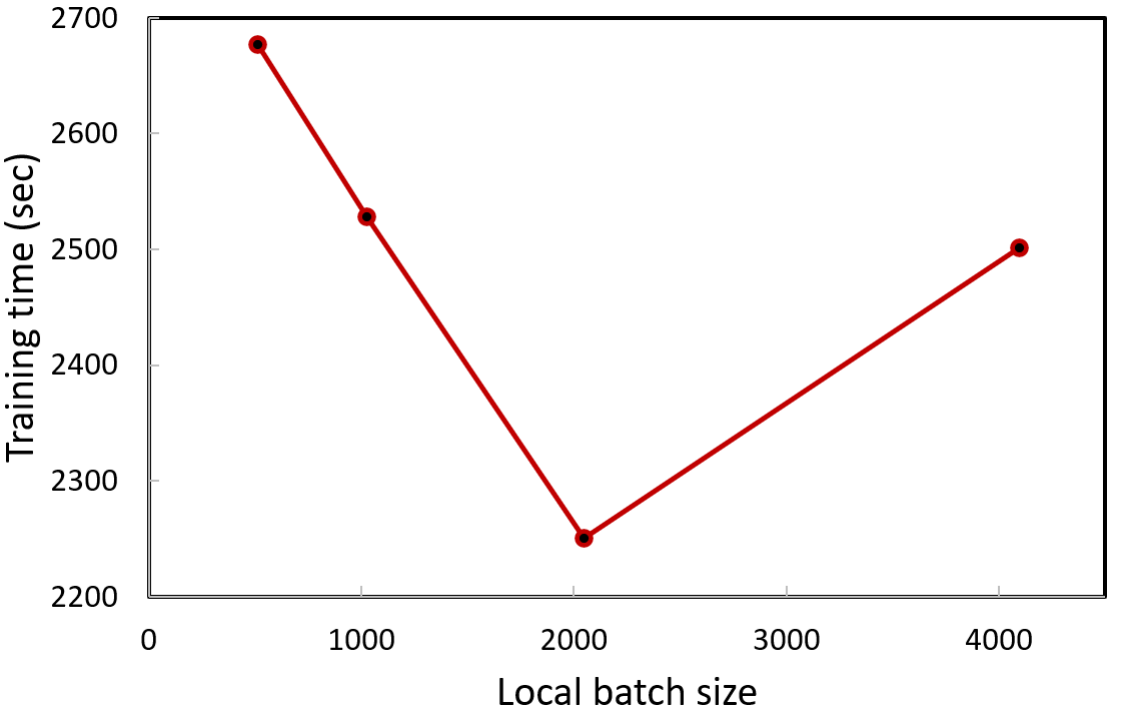}

\caption{
    Comparison of training time across using different local batch sizes of low-resolution data in the pretraining phase on Neuron Inverter. 
}
\label{chart:ni_bs}
\end{figure}
In the pretraining phase, the smaller sample size of low-resolution data allows a larger local batch size to be used.
In one-fusion CosmoFlow training, due to the short average epoch time of 1.76 seconds, further increasing the batch size leads to an almost unaffected pretraining time.
Therefore, in this section, we study the impact of using different local batch sizes in low-resolution pretraining with one-fusion MRMF on Neuron Inverter.
In original Neuron Inverter training, $512$ is used as the local batch size.
Thus, we increase the local batch size of low-resolution data from 512 to 4096 with the number of GPUs fixed.

Figure~\ref{chart:ni_bs} shows the end-to-end training of using one-fusion MRMF on the Neuron Inverter when using local batch sizes as 512, 1024, 2048, and 4096.
The same model accuracy convergence condition, 0.0400 validation loss, is used when using four local batch sizes.
We observe that the end-to-end training time shows a clear timing reduction when increasing the local batch size on low-resolution data before reaching a batch size of $4096$.
Given the number of training samples and GPUs remains constant, increasing the local batch size in turn increases the global batch size by a factor of the number of GPU devices. A larger global batch size in turn decreases the number of iterations per epoch, reducing the end-to-end training time.
However, increasing the global batch size may affect the end model accuracy due to the generalization gap.
As expected, using the local batch size of $4096$ on low-resolution data increases the total training time from $2250.21$ seconds when using $2048$ as the local batch size to $2501.51$ seconds.
These experiments indicate that increasing the local batch size in low-resolution data training could further reduce the training time, but excessively large sizes may hinder performance by delaying convergence.

\section{Conclusion}
\label{sec:conclusion}
In this paper, we propose a multi-resolution model fusion training method that pretrains models on multi-resolution data to help reduce the original problem's training time. 
Given a scientific dataset could be discretized into different resolutions, our fast training method can be applied to accelerate the end-to-end model training.
Our experiment results on two real-world scientific applications demonstrate that the proposed approach does not affect the end model accuracy while significantly reducing the computation costs.
We also empirically prove that negligible preprocessing time is incurred in generating multi-resolution data.
Although our study focused on the performance of applications involving 3D cosmology and 1D neuron data, the MRMF we preprosed is adaptable to a variety of data types and applications across different fields.
Considering that fusing models pretrained at different resolutions could reduce the time required for finetuning, generalizing this approach to Graph Neural Networks (GNNs) and other model architectures would be an intriguing avenue for future research.
\section{Acknowledgements}
This material is based upon work supported by the U.S. Department of Energy, Office of Science, Office of Advanced Scientific Computing Research, Scientific Discovery through Advanced Computing (SciDAC) program under Award Numbers DE-SC0021399.
This research used resources of the National Energy Research Scientific Computing Center (NERSC), a DOE Office of Science User Facility supported by the Office of Science of the U.S. Department of Energy under Contract No. DE-AC02-05CH11231 using NERSC award ASCR-ERCAP0028620.
This work is partially supported by the National Institute of Standards and Technology award number 70NANB19H005. This work is also supported by the National Science Foundation Graduate Research Fellowship under Grant No. DGE-2039655.
This work was supported by the Office of Advanced Scientific Computing Research, Office of Science, of the U.S. Department of Energy under Contract No. DE-AC02-05CH11231.

\bibliographystyle{elsarticle-num} 
\bibliography{mybib}
\end{document}